\documentclass[pra,twocolumn,showpacs,superscriptaddress]{revtex4}

\usepackage{graphicx}
\usepackage{amsmath}

\begin{document}


\title{FFLO Vortex Lattice States in Cold Fermionic-Atom Systems}
\author{Y.-P. Shim}
\affiliation{Department of Physics,
University of Texas at Austin, Austin, Texas 78712}
\author{R. A. Duine}
\affiliation{Department of Physics,
University of Texas at Austin, Austin, Texas 78712}
\affiliation{Institute for Theoretical Physics,
Utrecht University, Leuvenlaan 4, 3584 CE Utrecht, The Netherlands}
\author{A. H. MacDonald}
\affiliation{Department of Physics,
University of Texas at Austin, Austin, Texas 78712}
\date{\today}

\begin{abstract}
Condensation of atom pairs with finite total momentum
is expected in a portion of the phase diagram of a two-component
fermionic cold-atom system. This unusual condensate can be
identified by detecting the exotic higher Landau level(HLL)
vortex lattice states it can form when rotated.
With this motivation, we have solved the linearized gap equations
of a polarized cold atom system in a Landau level basis
to predict experimental circumstances under which HLL vortex lattice states occur.
\end{abstract}


\maketitle


\section{Introduction}

Polarized two-component fermion systems tend toward
finite pair momentum condensates because of the Fermi radius
mismatch between majority and minority components.
In superconductors, electron spin-polarization
can be induced by the application of an external field or by
proximity coupling to a ferromagnet.  Finite-momentum
Cooper pair condensates in spin-polarized superconductors,
Fulde-Ferrell-Larkin-Ovchinnikov (FFLO) states, were first proposed in the
early 1960's \cite{FF,LO}. One important consequence of finite-momentum
pairing in an isolated superconductor is a spatially inhomogeneous order
parameter. There have been many efforts in various
solid state systems to detect this exotic state, including recent
ones \cite{radovan2003,kakuyanagi2005}, but its definitive identification
has remained elusive.  The disorder that is inevitably present in a solid
state system may have played a role in the absence of a conclusive FFLO state
identification in studies of spin-polarized superconductors.

Experimental progress \cite{Zwierlein1,Zwierlein2,Martin,Hulet}
with fermionic cold-atom systems has given rise to a new strategy
for realizing  the FFLO state or the related Sarma state \cite{Sarma}
and has stimulated a great deal of theoretical
activity \cite{Radzihovsky,Torma,Chevy,Pieri,Duan,
Erich,Masud,Ho,Demler,Machida,Jani,Levin,Simons,Bulgac,gubbels2006}.
The tunability of the interaction between atoms via a
Feshbach resonance \cite{stwalley1976,tiesinga1993} has made it
possible to increase the strength of fermion pairing and has even
made the BEC-BCS crossover \cite{eagles1969,leggett1980,nozieres1985}
experimentally accessible. On the Bose-Einstein condensate (BEC) side of a
Feshbach resonance fermionic atoms form bosonic molecules which
condense at low temperatures. On the Bardeen-Cooper-Schrieffer
(BCS) side, the effective attractive interaction between fermion
atoms leads to BCS-type pairing. In between lies the so-called
unitarity limit \cite{burovski2006} in which no weakly-interacting
particle description applies.

Easy control over the population of two hyperfine states in a
trapped atom cloud, makes cold-atom systems a promising candidate
for FFLO state realization.  The FFLO state
competes \cite{Radzihovsky,Torma,Chevy,Pieri,Duan,
Erich,Masud,Ho,Demler,Machida,Jani,Levin,Simons,Bulgac,gubbels2006}
with a number of other states, including in cold atom systems
states with phase separated regions that are respectively
unpolarized and unpaired. The FFLO state is expected to occur on
the BCS side of the BEC-BCS crossover, at temperatures and
pressures close to the normal/superfluid phase boundary.
Population imbalance in cold atoms plays essentially the same role
as a Zeeman or exchange field in a superconductors since pairing
is dependent on energy measured from the Fermi energy for each
species of fermion. In both cases the Fermi radius of the majority
species exceeds the Fermi radius of the minority species and pairs
at the Fermi energy necessarily have non-zero total momentum.

One of the most obvious signatures of superfluidity in fermionic
cold-atom systems is the appearance of vortices and vortex
lattices when the system is rotated \cite{feder2004}. Indeed
recent experiments \cite{Zwierlein1} have observed vortex-lattice
structures in fermionic cold-atom systems close to the BEC-BCS
crossover region.  For this reason an obvious potential signature
of an FFLO state is the appearance of the exotic vortex-lattice
structures they are expected to form \cite{shimahara1997,klein2004,Yang}.
FFLO vortex lattices can be
wildly different from the usual hexagonal Abrikosov vortex
lattice. The structure of the vortex lattice is determined
mainly \cite{shimahara1997,klein2004,Yang} by the Landau level
index of its condensed fermion pairs; the Abrikosov lattice forms
when the Landau level index $j=0$, which is the closest
approximation to zero-total-momentum pairing allowed in a system
that has come to equilibrium in a rotating frame.  FFLO states in
the absence of rotation can imply $j>0$ Fermion pair condensation
in rotated systems. Vortices have been observed in systems with
population imbalance \cite{Zwierlein2}, but so far no unusual
vortex structures have been observed. (This could be due to
the fact that these experiments realize the gapless Sarma phase \cite{gubbels2006}
and another reason could be that the FFLO state is predicted
by weak-coupling theory while all experiments are in the unitary limit.)

With this motivation, we report on a study of the polarization
and interaction strength regime over which non-zero $j$ pairing is expected
in a rotating two-component Fermion system.  We consider only the BCS side
of the Feshbach resonance, on which FFLO physics occurs.
We consider three-dimensional systems for the sake of
definiteness, although two-dimensional systems could also be interesting
experimentally.  Working in the co-rotating reference frame, rotation
is equivalent to an external magnetic field and a reduction in radial
confinement strength.  All our explicit calculations are for a uniform
three-dimensional system and do not account for confinement.
In typical experiments the atomic Landau level splitting,
equal to $2 \hbar \Omega$ where $\Omega$ is the rotation frequency, is
much smaller than the Fermi energy.  In this limit the Landau level index
of the condensate could be determined by finding the optimal pairing
wavevector on the BCS superfluid/normal phase boundary in the absence
of rotation and using semiclassical quantization to add rotation to the
condensate effective action.
Here we use a fully quantum-mechanical approach, including Landau
quantization even at the level of the underlying unpaired fermions.
This approach is still relatively easy, partly because of the short-range
of the atom-atom attractive effective interaction, and has of the advantage
of determining the condensate Landau level index
more accurately, and allows us to comment on the rapid rotation regime which
might be approached experimentally in the future.
From now on we use the language
of the co-rotating frame so that the atoms experience an effective field
with cyclotron frequency $\Omega_c = 2 \Omega$.

\begin{figure}
    \includegraphics[width=0.7\linewidth]{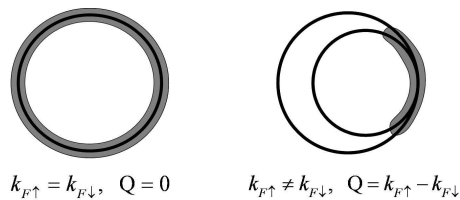}
    \caption{\label{fig:pairing}
             Low energy pairings for population balanced
             and unbalanced systems. Shaded regions indicate
             participating states for the low energy parings in $k$-space.
             $Q$ is the total momentum of the pairs,
             which is 0 for balanced systems and equal to
             the difference between Fermi wavevectors in unbalanced systems.
            }
\end{figure}

\begin{figure}
    \includegraphics[angle=0,width=0.65\linewidth]{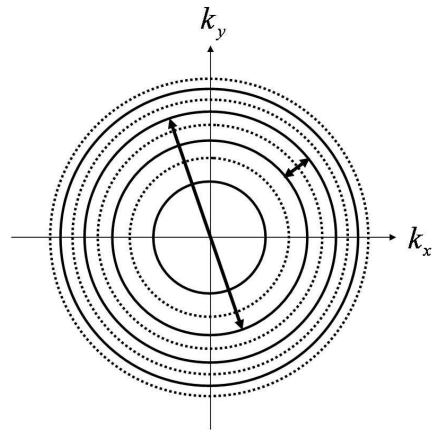}
    \caption{\label{fig:LL}
             Degeneracy of Landau levels.
             States between two dotted circles collapse
             into the solid circle. All the areas
             between two adjacent dotted circles are the same and
             solid circles have radii given by
             $\hbar^2 k^2/2m = \hbar \Omega_c (N+1/2)$.
             The arrows show the maximum and minimum momentum differences
             between particles in LL $N=1$ and $N=2$, which correspond
             qualitatively to the maximum and minimum of the COM momentum.
            }
\end{figure}

Pairing is most effective when the states to be paired are as
close to the Fermi energy as possible.  When there is no
population imbalance, pairs formed from electrons with opposite
momentum (zero total momentum) are abundant at low energies as
illustrated schematically in Fig.~\ref{fig:pairing}.
%
%
For unbalanced populations
the lowest energy pairs have total momentum equal to the
difference between Fermi wavevectors. In systems with an orbital
magnetic field linear momentum is not a good quantum number, but
the motion of a pair can still be separated into center-of-mass
and relative motion degrees-of-freedom.  In a magnetic field,
momentum space collapses into Landau levels whose degeneracy is
illustrated in Fig.~\ref{fig:LL} by partitioning of momentum space into
equal area segments centered on $\hbar \Omega_c (N+1/2)$.
%
%
A pair of electrons with given Landau level indices $N$ and $N'$ has finite
quantum amplitudes for all center of mass Landau level indices
from $0$ to $N+N'$ which correspond closely to the
distribution of center of mass (COM) kinetic energy values that
would be obtained by averaging over the corresponding regions of
momentum space illustrated in Fig.~\ref{fig:LL}. These quantum probability
amplitudes are the key ingredient in the linearized gap equations
discussed below. We derive linearized gap equations which
implicitly define the critical temperature for a phase transition
from the normal to the superfluid state for each COM LL and
determine the phase boundaries in parameter space. If excited COM
LL's  have a higher critical temperature than the lowest-lying COM
LL, this signals the occurrence of exotic vortex lattice states and
of FFLO states in the unrotated system.
%
%
\begin{figure}
    \includegraphics[angle=0,width=0.7\linewidth]{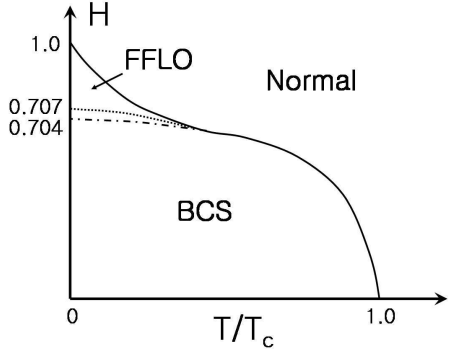}
    \caption{\label{fig:FFLO}
             BCS theory phase diagram for FFLO and BCS states as calculated,
             for example, in Ref.[~\onlinecite{Yang1}].
             Here, $H$ is the ratio of the Zeeman energy
             (or normal state chemical potential
             difference) to the zero-field energy gap.
             The dotted line marks the Clogston limit
             where the energies of normal and
             the zero-pairing momentum BCS state are identical.
             The FFLO state occurs near
             the boundary between normal and BCS states. }
\end{figure}
In Fig.~\ref{fig:FFLO}
the phase diagram is shown for a non-rotating homogeneous system. The maximum
value of the exchange field (or difference between normal state
chemical potentials) for which pairing still occurs is given
approximately by $H=\Delta_0/\sqrt{2}$, where $\Delta_0$ is the
BCS gap parameter at zero exchange field and zero temperature.
Beyond this so-called Clogston limit \cite{Clogston} the BCS state
is no longer stable.  The FFLO state is expected to occur in this
region of the phase diagram.

The rest of the paper is organized as follows.
In Sec.~\ref{sec:tceq} we derive COM Landau level index
dependent linearized gap equations for the critical
temperature of the rotating system. The numerical solution of this
equation is presented in Sec.~\ref{sec:numresults}. We finish in
Sec.~\ref{sec:discuss} with a discussion of our results, and present our
conclusions.   We postpone to this section a discussion of the
competition between phase separated states and FFLO states, which is
an issue for cold atoms but not for electrons in a solid because
of long-range repulsive Coulomb interactions.

\section{Linearized Gap Equations} \label{sec:tceq}

In this section we derive the linearized gap equation for
condensation of Fermion pairs with a definite COM Landau Level (LL) index.
We first consider the transformation between individual particle and
COM and relative states for two rotating atoms and then use
this to derive the gap equations, which
are implicit equations for the critical temperatures of
each COM LL index channel.

\subsection{Unitary Transformation}

To consider the pairing instability of a normal Fermi gas, we
first turn our attention to the description of scattering between
two atoms in a rotating reference frame.  The rotation is
represented by considering the atoms to be particles with unit charge
in an effective homogeneous orbital magnetic field.
The Hamiltonian for two particles is {\setlength{\arraycolsep}{1pt}
\begin{eqnarray}
\widehat{h}&=& \frac{1}{2m}
               \left(
                      -i\hbar \boldsymbol{\nabla}_{\mathbf{r}_1}
                     - \mathbf{A}(\mathbf{r}_1)
               \right)^2
             + \frac{1}{2m}
               \left(
                      -i\hbar \boldsymbol{\nabla}_{\mathbf{r}_2}
                     - \mathbf{A}(\mathbf{r}_2)
               \right)^2\nonumber\\
           &=& \frac{1}{2M}
               \left(
                     -i\hbar \boldsymbol{\nabla}_{\mathbf{R}}
                     -2 \mathbf{A}(\mathbf{R})
               \right)^2
             + \frac{1}{2\mu}
               \left(
                     -i\hbar \boldsymbol{\nabla}_{\mathbf{r}}
                     -\frac{\mathbf{A}(\mathbf{r})}{2}
               \right)^2~,\nonumber\\
\end{eqnarray}}
where $M=2m$,~$\mu=m/2$,~$\mathbf{R}=(\mathbf{r}_1 + \mathbf{r}_2)/2$ and
$\mathbf{r}=\mathbf{r}_1 - \mathbf{r}_2$. The vector potential
$\mathbf{A}(\mathbf{r})$ is defined by
$\boldsymbol{\nabla}\times\mathbf{A}(\mathbf{r})=
2m\Omega\,\hat{\mathbf{z}}$ where $\Omega$ is the angular rotation
frequency of the system and we assume that the rotation is around
the $z$-axis. In the Landau gauge,
$\mathbf{A}(\mathbf{r})=(0,2m\Omega x,0)$ and the individual atom
eigenfunctions with eigenvalues $\hbar \Omega (2 N + 1)$ are given
by
\begin{eqnarray}
&& \psi_{N,k_{i,y},k_{i,z}}(\mathbf{r}_{i})
  = \langle \mathbf{r}_{i} | N,k_{i,y},k_{i,z} \rangle \nonumber\\
&& \qquad = e^{i(k_{i,y}y_i+k_{i,z}z_i)}
            \phi_N(x_i + k_{i,y}l_B^2)/(L_y L_z)^{1/2} ~,\qquad
\end{eqnarray}
where $\phi_N(\mathbf{r})$ is the one-dimensional harmonic
oscillator eigenfunction and the effective magnetic length $l_B$
is defined by $\hbar^2/m l_B^2 = 2\hbar\Omega$. The eigenfunctions
are labeled by the momenta in $y$ and $z$ directions, and by the
LL index $N$. The eigenfunctions for the COM and relative
coordinates are the same, except that the effective magnetic
lengths are now $l_R = l_B / \sqrt{2}$ and $l_r = \sqrt{2} l_B$.
In terms of ladder operators,
\begin{equation}
\widehat{h}=\hbar\Omega_c(a_1^{\dag} a_1 + a_2^{\dag} a_2 +1)
= \hbar\Omega_c(a_R^{\dag} a_R + a_r^{\dag} a_r +1) ~,
\end{equation}
where $a_i=(l_B/\sqrt{2}\hbar)(\pi_{i,x}-i\pi_{i,y})$,
$\boldsymbol{\pi}_i=i\hbar \boldsymbol{\nabla}_i - \mathbf{A}(\mathbf{r}_i)$,
$a_R=(a_1+a_2)/\sqrt{2}$, $a_r=(a_1-a_2)/\sqrt{2}$,
and $\hbar \Omega_c = \hbar^2/m l_B^2 = 2\hbar\Omega$. The ladder operators
can then be used to derive \cite{AHM1} an explicit expression
for the unitary transformation
between individual particle and COM and relative two-atom states:
\begin{eqnarray}
&& \langle \mathbf{r_1},\mathbf{r_2}|N,k_{1,y},k_{1,z};M,k_{2,y},k_{2,z}
   \rangle \nonumber\\
&&  = \sum_{j=0}^{N+M} B_j^{NM} \langle \mathbf{R},\mathbf{r}| j,
K_y, K_z ; N+M-j, k_y, k_z \rangle ~,\qquad
\end{eqnarray}
where
\begin{eqnarray*}
&& K_y = k_{1,y}+k_{2,y}~,~K_z = k_{1,z}+k_{2,z}~, \\
&& k_y = (k_{1,y}-k_{2,y})/2~,~k_z = (k_{1,z}-k_{2,z})/2~,
\end{eqnarray*}
and
\begin{eqnarray}
B_j^{NM} &=& \left[
                    \frac{j!(N+M-j)!N!M!}{2^{N+M}}
              \right]^{1/2} \nonumber\\
         &\times&
              \sum_{m=0}^{j}
              \frac{(-)^{M-m}}{(j-m)!(N+m-j)!(M-m)!m!}~.\qquad
\end{eqnarray}
%
%
\begin{figure}
    \includegraphics[angle=0,width=0.9\linewidth]{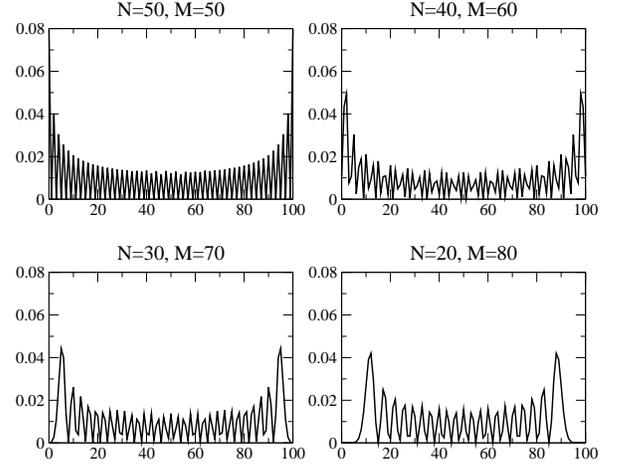}
    \caption{\label{fig:BjNM}
             $|B_j^{NM}|^2$ vs $j$ with $N+M=100$ for different $N$'s.
             The horizontal axes are $j$ and the vertical axes are
             $|B_j^{NM}|^2$.}
\end{figure}
It follows that $B_j^{NM}$ is the probability amplitude for two
atoms in LLs $N$ and $M$, respectively to have COM LL
$j$ and the relative motion LL $N+M-j$. When $N=M$,
$|B_j^{NM}|^2$ has maxima for $j=0$ and $j=N+M$. However, if $N
\neq M$, $|B_j^{NM}|^2$ can have a maximum for intermediate $j$,
which means that for two atoms in different LLs, the most probable
COM LL can be different from zero or $N+M$ as shown in Fig.~\ref{fig:BjNM}.
The smooth envelope apparent in these figures is simply the zero-field
probability distribution of the COM kinetic energies given the Fermi
momenta of two individual particles. The COM energy is maximum for
parallel momentum and minimum for oppositely oriented individual
particle momenta. This coefficient plays
an important role in determining the pairing COM LL in condensed
states.

\subsection{Bethe-Salpeter Equations}

\begin{figure}
    \includegraphics[angle=0,width=0.9\linewidth]{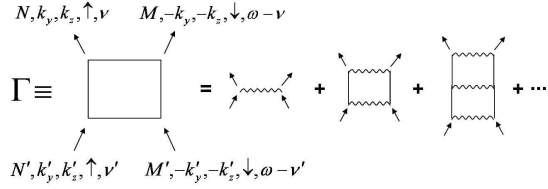}
    \caption{\label{fig:ladder}
             Ladder diagrams to be summed for scattering function $\Gamma$.
            }
\end{figure}

The pairing instability in a Fermi gas is signaled by a divergence
of the many-body scattering function \cite{schriefferbook}, which
we approximate using the Bethe-Salpeter equation summarized by the
finite-temperature Feynman diagrams illustrated in
Fig.~\ref{fig:ladder}.
We consider a system consisting of two
hyperfine species denoted by $\uparrow$ and $\downarrow$. For
definiteness we assume that the two species have the same energy
spectrum but allow for different densities and therefore different
chemical potentials. Population imbalance is relatively easy to
achieve experimentally and the life-time of each hyperfine state
is long enough compared to experimental time scales to justify the
use of equilibrium statistical mechanics with separate particle
reservoirs for the two species. The many-body scattering function
is calculated by summing the ladder diagrams \cite{Stoof,AHM1}(see
Fig.~\ref{fig:ladder}).
%
%
Generalizing the calculations of
Ref.[~\onlinecite{AHM1}] to three dimensions from two and we find that
the total two-particle scattering
function can be written as a sum over different COM Landau level
index channels:
\begin{eqnarray}
\label{eq:totalscattfunc}
&& \Gamma(N,M,k_y,k_z ; N',M',k_y',k_z' ;i\omega) \nonumber\\
&&  =\sum_j B_j^{NM} B_j^{N'M'} \gamma_j(N,M,k_y,k_z ,
N',M',k_y',k_z' ;i\omega).\quad
\end{eqnarray}
where the partial scattering function for COM LL $j$
\begin{widetext}
\begin{eqnarray}
\label{eq:partialscattfunc}
   \gamma_j(N,M,k_y,k_z ; N',M',k_y',k_z' ;i\omega)
                &=& \langle N+M-j,k_y,k_z | \widehat{V} | N'+M'-j,k'_y,k'_z
                    \rangle \nonumber\\
   &+& \sum_{N'',M''} \sum_{ k''_y,k''_z }  \left|B_j^{N''M''}\right|^2
       \langle N+M-j,k_y,k_z | \widehat{V} | N''+M''-j,k''_y,k''_z
       \rangle \nonumber\\
   && \qquad \times ~ K_{N'',M'',k''_z}(i\omega)
             \gamma_j(N'',M'',k''_y,k''_z ; N',M',k_y',k_z'.
             ;i\omega)~.
\end{eqnarray}
\end{widetext}
In Eq.~\eqref{eq:partialscattfunc}
\begin{eqnarray}
K_{N,M,k_z}(i\omega) &=&
\frac{1-f(\xi_{N,k_z,\uparrow})-f(\xi_{M,-k_z,\downarrow})}
     {i\hbar\omega - \xi_{N,k_z,\uparrow} - \xi_{M,-k_z,\downarrow}}~;\\
\xi_{N,k_z,\sigma} &=& \varepsilon_{N,k_z} -\mu_{\sigma}~;\\
\varepsilon_{N,k_z} &=& \hbar
\Omega_c\left(N+\frac{1}{2}\right)+\frac{\hbar^2 k_z^2}{2m}~,
\end{eqnarray}
and $f(\xi)$ is the Fermi distribution function. In the case of a
delta-function interaction $V(\mathbf{r})=-V_0 \delta(\mathbf{r})$
we have that
\begin{eqnarray}\label{V}
\langle N+M-j,k_y,k_z | \widehat{V} | N'+M'-j,k'_y,k'_z \rangle
                                  \quad\quad\quad && \nonumber\\
= -V_0 \phi^r_{N+M-j}(k_y l_r^2) \phi^r_{N'+M'-j}(k'_y l_r^2)
(1/L_y L_z)  ~,\quad\quad
\end{eqnarray}
where $\phi^r_N$ is the one-dimensional harmonic oscillator
eigenfunction in relative coordinates. Using this property and the
orthogonality of the relative motion harmonic oscillator
wavefunctions we find that
{\setlength{\arraycolsep}{1pt}
\begin{eqnarray}\label{gamma_j}
&& \gamma_j(N,M,k_y,k_z ; N',M',k_y',k_z' ;i\omega) \qquad\nonumber\\
&=& \frac{-V_0}{L_y L_z} \phi^r_{N+M-j}(k_y l_r^2) \phi^r_{N'+M'-j}(k'_y l_r^2)
    \qquad\nonumber\\
&\times&  \left(1+\frac{V_0}{4\pi l_B^2 L_z}
                  \sum_{N'',M'',k''_z} K_{N'',M'',k''_z}(i\omega)
                  \left|B_j^{N''M''}\right|^2
          \right)^{-1}~. \nonumber\\
\end{eqnarray}}

\subsection{$T_{c}$ Equation}

As mentioned before, the instability of the normal state due to
pairing is signaled by the divergence of the many body scattering
function $\Gamma(i\omega=0)$, and therefore a diverging
$\gamma_j(i\omega=0)$ means that pairs with COM LL $j$ are
unstable to condensation.  This instability condition for the
scattering function is equivalent to the linearized gap equation
which defines the critical temperature \cite{schriefferbook} in
mean-field theory. (In the mean-field-theory for the ordered state \cite{Akera}
the order parameter can be expressed in terms of partial
contributions from each COM LL channel.  When the order parameter
is small the various channels decouple and the partial
contribution from a given channel vanishes at the same point at
which the normal state partial scattering function diverges.) From
Eq.~\eqref{gamma_j}, we get an implicit equation for the
critical temperature $T_{c}^j$ for each COM LL $j$, which
reads
\begin{equation}
\frac{1}{V_0} = \frac{1}{4\pi l_B^2 L_z}
                \sum_{N,M,k_z}
                \frac{1-f(\xi_{N,k_z,\uparrow})-f(\xi_{M,-k_z,\downarrow})}
                     {\xi_{N,k_z,\uparrow} + \xi_{M,-k_z,\downarrow}}
                \left|B_j^{NM}\right|^2~.
\end{equation}

Unlike the BCS superconductors, for which retarded phonon-mediated
attractive interactions have a natural ultraviolet cut-off, there
is no cut-off in this equation and the summation is over all
states. Hence, as it stands, this equation diverges, because of
the assumption of a $\delta$-function interaction. To remove this
divergence, we need to recognize that the true atom-atom
interaction is short-ranged compared to relevant atomic
wavelengths but not a $\delta$-function.
Using the exact relation between scattering length and interaction
strength (see Eq.~\eqref{eq:scatlength} in the appendix)
we remove the interaction strength $V_0$ by renormalizing to the
scattering length \cite{Stoof} in the $T_{c}^j$ equation and obtain convergent
sums over intermediate states. The equation for $T_{c}^j$ then
becomes
\begin{widetext}
\begin{equation}\label{Tc}
-\frac{1}{k_{F0} a_{\rm sc}}
 = \frac{\hbar \Omega_c}{2 \pi k_{F0}} \sum_{N,M} \int dk_z
   \left[
         \frac{1-f(\xi_{N,k_z,\uparrow})-f(\xi_{M,-k_z,\downarrow})}
              {\xi_{N,k_z,\uparrow} + \xi_{M,-k_z,\downarrow} }
              \left|B_j^{NM}\right|^2
        -\frac{1}{\varepsilon_{N,k_z}+\varepsilon_{M,-k_z}}
              \left|B_0^{NM}\right|^2
   \right]~,
\end{equation}
\end{widetext}
where $k_{F0}$ is the Fermi wavevector of the unpolarized system
without rotation. The left-hand side of Eq.~\eqref{Tc} is
experimentally measurable.  We determine $T_{c}$ as a function of
$1/k_{F0}a_{\rm sc}$ by solving this implicit equation combined
with implicit equations for the temperature-dependent chemical
potentials $\mu_\sigma$
\begin{equation}
n_{\sigma} = \frac{1}{V} \sum_{N,k_y,k_z} f(\varepsilon_{N,k_z} -
\mu_{\sigma})~,
\end{equation}
where $n_\sigma$ is the density of atoms in hyperfine state
$\sigma$, and $V$ is the total volume of the system. In the next
section we present numerical results obtained by solving these equations.

\section{Numerical Results}\label{sec:numresults}

We calculate $T_{c}^j$ for each COM LL $j$ for various rotation
frequencies, interaction strengths and polarizations and in this
way determine the phase boundaries in the parameter space spanned
by $\hbar\Omega_c$, $a_{\rm sc}$ and the polarization. We fix the
total density of the system $n_{\rm tot}$ and used the
polarization $p$ as a parameter. The polarization is defined by
\begin{equation}
p=\frac{n_{\uparrow}-n_{\downarrow}}{n_{\uparrow}+n_{\downarrow}}~,
\end{equation}
where $n_{\uparrow}$ is the density of the majority species and
$n_{\downarrow}$ is the density of the minority species. Hence,
the density of atoms in species $\sigma$ [$\sigma=+1$ ($-1$)
corresponds to $\uparrow$ ($\downarrow$)] is given by
\begin{equation}
n_{\sigma} = \frac{1 + \sigma p}{2} \cdot n_{\rm tot}~.
\end{equation}
%
%
\begin{figure}
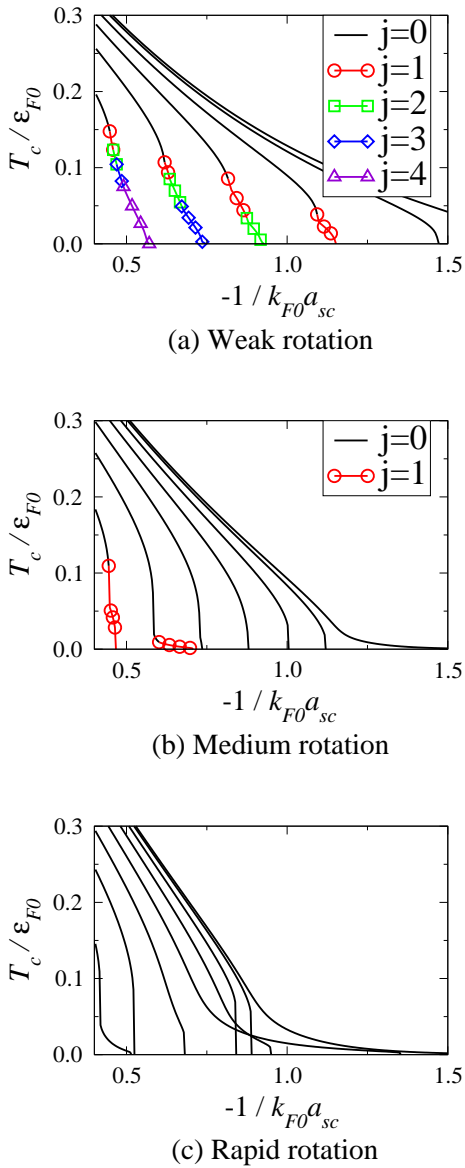

    \includegraphics[width=0.7\linewidth]{plot2_0.02.eps}
    \\[20pt]
    \includegraphics[width=0.7\linewidth]{plot2_0.17.eps}
    \\[20pt]
    \includegraphics[width=0.7\linewidth]{plot2_0.50.eps}
    \\[20pt]
    \caption{ \label{fig:plot2}
                (Color online) Critical $T_{c}$ vs $-1/k_{F0}a_{\rm sc}$.
                (a) $\hbar \Omega_c /\varepsilon_{F0}=0.02$.
                The curves are for different polarizations
                0.0, 0.1, 0.2, 0.3, 0.4, 0.5 from top to bottom.
                (b) $\hbar \Omega_c /\varepsilon_{F0}=0.17$.
                Polarizations are from 0.0 to 0.6.
                (c) $\hbar \Omega_c /\varepsilon_{F0}=0.50$.
                Polarizations are from 0.0 to 0.7.
               }
\end{figure}

The relationship between $T_{c}^j$ and interaction strength is
illustrated in Fig.~\ref{fig:plot2}. The true critical temperature
for the system is the largest value of $T_{c}^j$.
\begin{equation}
  T_{c} = \max \left\{ T_{c}^j \right\}~.
\end{equation}
At weak rotation [Fig.~\ref{fig:plot2}~(a)], the transition
temperature $T_{c}$ for zero polarization shows the usual
behavior \cite{Gorkov} $T_{c} \propto {\rm exp}(-1/k_{F0}a_{\rm
sc})$ and the highest $T_{c}^j$ is for the $j=0$ channel
regardless of the interaction strength.  In this circumstance we
expect the
system will have a standard Abrikosov vortex lattice.
The critical temperature decreases as
polarization increases and superfluidity is suppressed above some
critical polarization. It is more easily suppressed at weak
interaction. FFLO states, which correspond to nonzero $j$, occur
at strong interaction and high polarization.
We emphasize that these
states will have very distinct \cite{Yang} vortex lattices, more open than the
hexagonal Abrikosov lattices and qualitatively different for each value of $j$.
It should be quite obvious experimentally when a $j \ne 0$ vortex lattice
occurs. We caution, however, that as the temperature drops below the critical
temperature, different values of $j$ will mix in the condensate \cite{Akera,AHM1},
the $j=0$ component will grow in weight even if it doesn't have the maximum $T_c$.
We speculate that the phase transition between finite momentum FFLO states and
zero-momentum BCS states, which occurs at zero field, is replaced in a field by
a smooth crossover between open and close-packed hexagonal lattices.
The best place to search experimentally for an exotic vortex lattice
is close to the superfluid/normal phase boundary
as possible by varying either temperature or interaction strength.
Indeed it appears advisable to conduct experiments in
systems with the smallest order parameter strength
for which it is possible to reliably visualize the vortex lattice.
Both the relatively large polarizations and strong interactions
required for the appearance of $j \ne 0$ solutions,
and the ability to tune parameters over wide ranges in atomic systems,
demonstrate the exceptional potential of tunable cold atom systems
in the hunt for FFLO vortex latices.
The greatest obstacle to realization of the FFLO state is
likely competition with phase separated
states. We return to this point again later.

The results reported in Fig.~\ref{fig:plot2}~(a) can be understood
qualitatively using quite simple considerations.  When the
temperature is low, weak pairing is expected to be dominated by
states at the Fermi energy.  For that reason, the zero-field
pairing wavevector on the phase boundary is expected to be close
to $k_{F\uparrow}-k_{F\downarrow}$ when $T_{c} \to 0$, {\em i.e.}
when the interactions are just strong enough to cause pairing.
Using a small $p$ approximation it follows that the pairing
wavevector for $T_{c} \to 0$ is given approximately by
\begin{equation}
Q = \frac{ 2 k_{F0}p}{3} .
\end{equation}
The Landau level index at finite fields can be estimated
by quantizing the pairing wavevector. This gives
\begin{equation}
j \approx \frac{(\hbar^2 Q^2)/4m}{\hbar \Omega_c}
\simeq \frac{\varepsilon_{F0}}{\hbar \Omega_c} \frac{2 p^2}{9}.
\label{eq:jestimate}
\end{equation}
It is easy to check that this equation is quite consistent with
the numerical results we have obtained. For smaller values of
$\hbar \Omega_c$ we therefore are confident that even larger
values of $j$ should occur, although exotic vortex lattice may
again be confined even more strongly to the region close to the
phase boundary. For a given value of polarization, the value of
$j$ decreases with increasing interaction strength because $T_{c}$
moves to higher temperatures, reemphasizing the importance of
pairing precisely at the Fermi energy.

Fig.~\ref{fig:plot2}~(b)~and~(c) show results for systems with
larger values of $\hbar \Omega_c$ than have been reported in
experiments to date. One observation is that non-zero $j$ states
are less likely to occur at large $\hbar \Omega_c$ and appear only
at very high polarization and strong interactions.  This property
is explained by Eq.~\eqref{eq:jestimate}. Indeed one can check
that the appearance of non-zero $j$ values is again consistent
with this estimate. Other new features that emerge in these
figures are due mainly to large LL quantization effects.  At very
high rotation frequency [Fig.~\ref{fig:plot2}~(c)], only the $j=0$
COM LL is realized.  Note that at high temperature, all the graphs
look similar.($T_{\rm c}$ decreases monotonically as the
polarization increases and as the interaction strength decreases.)
$T_{c}$ is more weakly dependent on the rotation frequency.
This is because the thermal energy is comparable to or larger than
the energy quantization due to rotation. On the other hand, at low
temperatures, the LL quantization effects become important because
the particles have one-dimensional densities-of-states for each
Landau level leading to peaks in pairing (at least in this
mean-field-theory calculation) when any Landau level is just
slightly occupied.  The non-monotonic density of states becomes
important when the LL spacing is much bigger than the temperature.
In this case, we expect non-monotonic behavior that is sensitive
to the density of both hyperfine species; we expect non-monotonic
dependence on polarization and the occasional appearance of strong
condensates at very large polarizations. Some of this
non-monotonic behavior is evident in Fig.~\ref{fig:plot2}~(c).
%
%
\begin{figure}
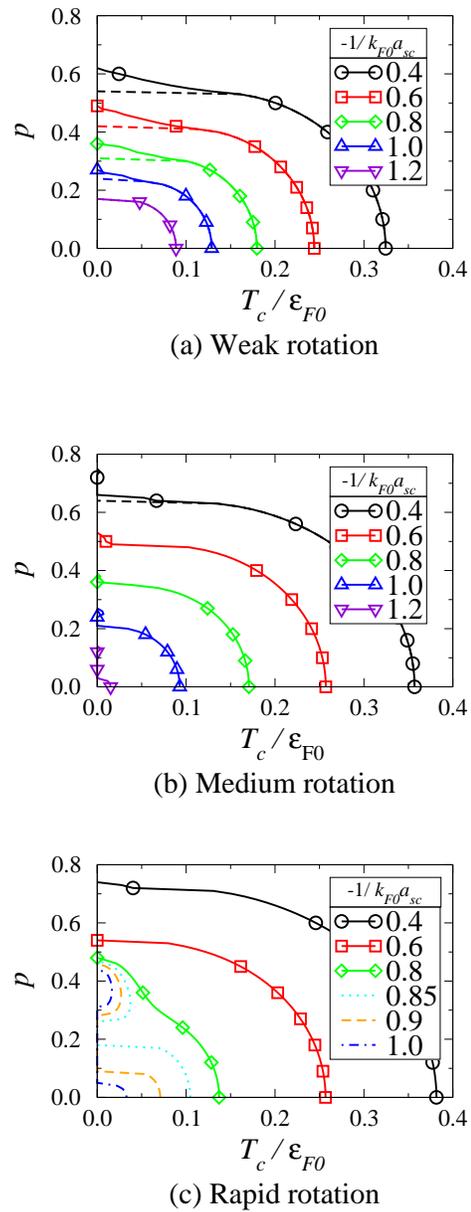

    \includegraphics[width=0.7\linewidth]{plot3_0.02.eps}%
    \\[20pt]
    \includegraphics[width=0.7\linewidth]{plot3_0.17.eps}%
    \\[20pt]
    \includegraphics[width=0.7\linewidth]{plot3_0.50.eps}%
    \\[20pt]
    \caption{\label{fig:plot3}
            (Color online) Polarization vs $T_{c}/\varepsilon_{F0}$.
            (a) $\hbar \Omega_c /\varepsilon_{F0}=0.02$.
            Curves are for different values of $-1/k_{F0}a_{\rm sc}$.
            (b) is for $\hbar \Omega_c /\varepsilon_{F0}=0.17$ and
            (c) is for $\hbar \Omega_c /\varepsilon_{F0}=0.50$.
            Dashed lines in (a) and (b) shows the $T_{c}$ curves for $j=0$
            and all the curves in (c) corresponds to $j=0$.
           }
\end{figure}

In Fig.~\ref{fig:plot3} we show the phase boundaries {\em vs.}
polarization and temperature for a series of interaction
strengths. For slow rotation [Fig.~\ref{fig:plot3}~(a)] it is
similar to the usual BCS-FFLO phase diagram (compare with
Fig.~\ref{fig:FFLO}). At higher rotation frequencies, shown in
Fig.~\ref{fig:plot3}~(b), FFLO states are less likely to occur.
The transition temperature still decreases monotonically as the
polarization increases and above some critical polarization, the
normal state prevails. At very high rotation frequencies, shown in
Fig.~\ref{fig:plot3}~(c), the LL quantization effects become more
important and we observe reemergence of condensed states at around
$p=0.4$. The difference of the Fermi energies at this polarization
is exactly equal to the LL spacing and the dominant pairing occurs
between individual particles whose Landau level indices differ by
one.
%
%
\begin{figure}
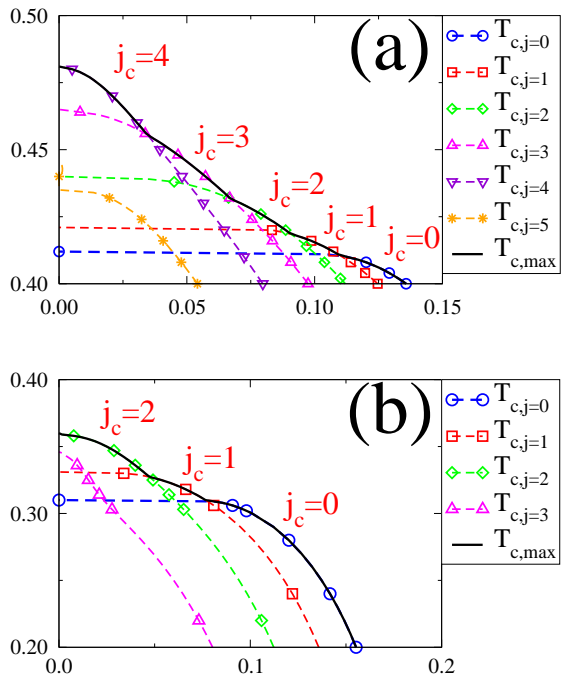

    \includegraphics[width=0.85\linewidth]{plot3_0.02_LHS0.6_large.eps}
    \\[20pt]
    \includegraphics[width=0.85\linewidth]{plot3_0.02_LHS0.8_large.eps}
    \\[20pt]
    \caption{\label{fig:plot3_large}
             (Color online) Enlarged figures of Fig.~\ref{fig:plot3}~(a)
             for $-1/k_{F0}a_{\rm sc}$= (a) 0.6 and (b) 0.8
             near the phase boundaries between FFLO states and normal fluid.
             The horizontal axis shows $T_{c}/\varepsilon_{F0}$
             and the vertical axis is polarization.
             We calculate $T_{c}$ for different $j$'s
             and determine the optimal $j$
             that gives the highest $T_{c}$.
            }
\end{figure}

In Fig.~\ref{fig:plot3_large} we show an enlargement of the
phase diagram for the FFLO state, showing also the critical
temperatures for a number of different COM LL index channels $j$
in addition to the one with the largest $T_c$. When the
polarization is small, $j=0$ pairing leads to the highest $T_{c}$;
that is $j=0$ is the optimal pairing channel for condensation
which we denote as $j_c$. As the polarization increases, $T_{c}^j$
for nonzero $j$ is larger than $T_{c}^{j=0}$ and $j_c$ increases
with the polarization. This is analogous to having an increasing
pairing COM momentum with increasing polarization field in the
zero-field case.  For a given value of $\hbar \Omega_c$, non-zero
values of $j_c$ are more likely when interactions are stronger,
because the superfluid has to be able to withstand the ill effects
of polarization out to a sufficiently large value of $p$. If the
interaction is too weak, no non-zero $j$ pairing can occur and
$j_c$ is zero.
%
%
\begin{figure}
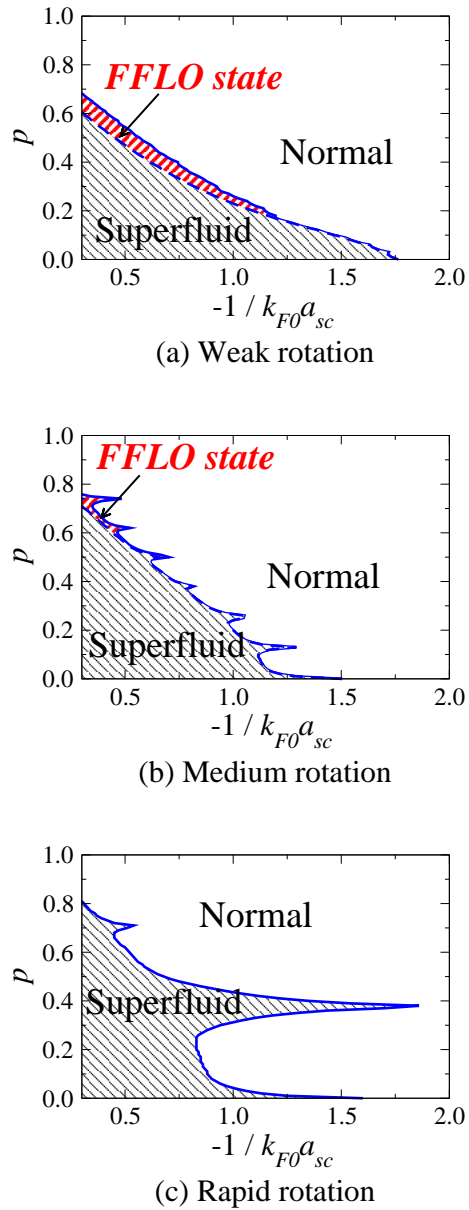

    \includegraphics[width=0.7\linewidth]{plot4_0.02.eps}%
    \\[20pt]
    \includegraphics[width=0.7\linewidth]{plot4_0.17.eps}%
    \\[20pt]
    \includegraphics[width=0.7\linewidth]{plot4_0.50.eps}%
    \\[20pt]
    \caption{\label{fig:plot4}
             (Color online) Polarization vs $-1/k_{F0}a_{\rm sc}$.
             We calculate $T_c$ for polarizations from 0 to 1
             with increment 0.01 and choose the largest one
             that has a finite $T_c$.
             (a) $\hbar \Omega_c /\varepsilon_{F0}=0.02$
             (b) $\hbar \Omega_c /\varepsilon_{F0}=0.17$ and
             (c) $\hbar \Omega_c /\varepsilon_{F0}=0.50$.
             Solid blue curves show phase boundary
             between normal fluid and superfluid and dashed blue curves in (a)
             and (b) show phase boundary for COM LL $j=0$.
            }
\end{figure}

In Fig.~\ref{fig:plot4} we plot the phase diagram {\em vs.}
polarization and effective interaction space for slow,
intermediate, and rapid rotations.  The critical polarization
decreases as the interaction strength decreases for weak rotations
[Fig.~\ref{fig:plot4}~(a)], as seen in
experiment \cite{Zwierlein2}. The regions labelled FFLO in this
figure have $j \ne 0$ condensates at the normal superfluid
boundary.  Quite generally this behavior occurs only in a small
region along the boundary between the superfluid and normal state
in the regime of large polarization and strong interactions.
Faster rotation generally suppresses FFLO states, as emphasized
earlier, but the superfluid phase can be realized at high
polarization and weak interaction by tuning the system such that the
Fermi energy mismatch between majority and minority species is an
integer times the LL spacing, and the Fermi energies are close to
a quantized LL energy. In Fig.~\ref{fig:plot4}~(c), we see that
big peaks occur if these conditions are met. At zero polarization
$\varepsilon_{F\uparrow}=\varepsilon_{F\downarrow}=1.96$, in units
of LL spacing, and the lowest LL is at 0.5. For $p=0.41$,
$\varepsilon_{F\uparrow}=2.52$ and
$\varepsilon_{F\downarrow}=1.52$ so that the Fermi energy
difference is exactly the LL spacing and each Fermi energy is very
close to the LLs.
For $p=0.72$, $\varepsilon_{F\uparrow}=2.80$ and
$\varepsilon_{F\downarrow}=0.80$.

\section{Discussion and Conclusions}\label{sec:discuss}

In summary, we have derived an equation for the superfluid
critical temperature in rotating Fermionic cold-atom systems,
incorporating Landau level quantization effects. Using this
equation we have calculated the phase boundary between the normal
and superfluid phase considering pairing in different
center-of-mass Landau levels. We find that states with higher
Landau level condensates can occur on the boundary between the
normal and superfluid phase regions in a parameter space that can
in principle be explored systematically by taking advantage of
Feshbach resonances and of the ability to create arbitrary degrees
of hyperfine state polarization in an atom cloud.  These FFLO
vortex lattice states will have distinct vortex
lattices \cite{shimahara1997,klein2004,Yang} which should aid their
identification.
High polarization and strong
interactions are required to realize the FFLO state. At high
rotation frequency, features that originate from rotational
quantization effects play an important role and we find that for
certain parameters the superfluid phase persists to high
polarization.

The regime where the FFLO
state occurs in rotating systems seems accessible to
experiment, and hence we believe that these exotic vortex
structures are observable. The greatest obstacle to their
observation may be competition with states in which the
atoms phase separate into regions with condensation but
no polarization and regions with polarization but no
condensation.  We believe that FFLO physics would almost
certainly occur if phase separation could be suppressed.
Phase separation does not occur for electrons in a
superconducting metal, and cannot because of the large
Coulomb energy price that would have to be paid. One
possibility for suppressing phase separation in atomic
systems with attractive interactions, is to artificially
create the necessary weak but long range repulsive interactions
by electrically inducing dipoles \cite{marinescu}
in a pancake shaped \cite{lewenstein},
but not necessarily quasi-two-dimensional trapped atom system.
The typical dipole-dipole interaction energy is
$p^2/R^3 \sim p^2 n \sim \alpha^2 E^2 n$
where $p$ is the dipole moment induced by the external electric field $E$,
$R$ is the average inter-atom distance, $n$ is the density of the atoms
and $\alpha$ is the polarizability of the atom. If this energy is
much smaller than the typical atom-atom interaction energy
$\varepsilon_{F0} (k_{F0} |a_{sc}|)$, then the physics on short length scales
does not change much.
On the other hand, if the energy cost of the whole system
due to the long range dipole interaction when the system is phase separated
is much larger than the condensation energy gain,
phase separation can be suppressed.
Thus, $p^2 n^2 V \gg D(0)\Delta_0^2 \sim N\varepsilon_{F0} e^{-\pi/k_{F0}|a_{sc}|}$
where $D(0)$ is the density of states at the Fermi level,
$N$ is the number of atoms and $V$ is the volume of the system.
These conditions lead to a condition for the external electric field
\begin{equation}
e^{-\pi/k_{F0}|a_{sc}|}
\ll \frac{\alpha^2 n E^2}{\varepsilon_{F0}} \ll
k_{F0}|a_{sc}|
\end{equation}
which can be easily satisfied for small $k_{F0}|a_{sc}|$.
FFLO states are most likely expected to occur near the critical temperature $T_c$
while experimentally observed phase-separated states are well below $T_c$.
It is known that phase separation is less likely at higher temperatures
so it could be possible to observe FFLO states near $T_c$ without explicitly
suppressing phase separation.

Finally we mention that peculiar additional interesting effects occur
because of Landau level
quantization if the rotation frequency is sufficiently large.
Very large rotation frequencies have been achieved in experiments with
bosonic atoms \cite{schweikhard2004}. We believe, therefore, that
there is no fundamental obstacle to approaching the
rapid-rotation limit with Fermions.  Although we have used
mean-field-theory here to study this regime, there is every reason
to expect unanticipated properties to emerge from strong quantum fluctuations
and correlations.  At sufficiently rapid rotations, it should be
possible to for the first time study the fractional quantum Hall
effect in fermion systems with attractive interactions \cite{Regnault}.

This work was supported by the National Science Foundation under
grant DMR-0606489 and by the Welch Foundation.


\begin{appendix}
\appendix*

\section{ Two-Body Transition Matrix and Scattering Length
          in Systems with Orbital Magnetic Field }

In this appendix, we derive the relation between the scattering length
and the strength of the delta-function like particle-particle interaction
in a system with orbital magnetic field.
The two-body transition operator for scattering at energy $z$ is
defined by
\begin{eqnarray}
\widehat{T}^{2B}(z) &\equiv& \widehat{V} + \widehat{V}
\frac{1}{z-\widehat{H}_0 } \widehat{V}
                           + \cdots \nonumber\\
&=& \widehat{V} + \widehat{V} \frac{1}{z-\widehat{H}_0 }
\widehat{T}^{2B}(z)~,
\end{eqnarray}
where $\widehat{V}$ is the particle-particle interaction and
$\widehat{H}_0$ is the non-interacting part of the two-body
hamiltonian. The matrix elements of this transition operator
satisfy the Lippman-Schwinger equation. Noting that scattering
conserves the COM motion, we calculate the $T$-matrix elements in
relative motion Hilbert space with COM LL $j$ and COM momenta
$K_y$ and $K_z$. Notice that the relative motion $T$-matrix
does not depend on the $y$-component of the total momentum $K_y$
in the Landau gauge. In the COM and relative coordinate representation
we find
\begin{widetext}
\begin{eqnarray}
&& \langle N+M-j,k_y,k_z |\; \widehat{T}^{2B}(j,K_z;z) \;
                         | N'+M'-j,k_y',k_z' \rangle  \nonumber\\
&& = \langle N+M-j,k_y,k_z | \widehat{V} | N'+M'-j,k_y',k_z'
                                         \rangle \qquad\qquad \nonumber\\
&& + \sum_{N'',M''}\sum_{k_y'',k_z''}
     \frac{ \langle N+M-j,k_y,k_z | \widehat{V} | N''+M''-j,k_y'',k_z'' \rangle
           |B_j^{N''M''}|^2
           \langle N''+M''-j,k_y'',k_z'' | \widehat{V} | N'+M'-j,k_y',k_z' \rangle
          }
          {z-\varepsilon_{N'',K_z/2+k_z''}
           -\varepsilon_{M'',K_z/2-k_z''}}
       \nonumber\\
&& + \cdots ~.
\end{eqnarray}
Using Eq.~\eqref{V} we have that,
\begin{eqnarray}
&& \langle N+M-j,k_y,k_z |\; \widehat{T}^{2B}(j,K_z;z) \;
                         | N'+M'-j,k_y',k_z' \rangle \nonumber\\
&& = \frac{-V_0}{L_y L_z}\phi^r_{N+M-j}(k_y l_r^2)\phi^r_{N'+M'-j}(k'_y l_r^2)
    \left[ 1 - \frac{-V_0}{4\pi l_B^2 L_z} \sum_{N'',M'',k_z''}
               \frac{|B_j^{N''M''}|^2}
                    {z-\varepsilon_{N'',K_z/2+k_z''}
                      -\varepsilon_{M'',K_z/2-k_z''}}
    \right]^{-1}~.
\end{eqnarray}
\end{widetext}
For a dilute atomic gas, all the relevant energies are small
compared to $\hbar^2/m r_V^2$ where $r_V$ is the interaction range.
We are therefore allowed to neglect the energy dependence of the two-body
$T$-matrix \cite{Stoof}. (Note that the energy does not depend on
$k_y$.) Hence we have that
\begin{eqnarray}
\label{eq:appendix}
&& \langle N+M-j,k_y,k_z |\; \widehat{T}^{2B}(j,K_z;z) \;
                         | N'+M'-j,k_y',k_z' \rangle \nonumber\\
&& \approx \langle 0,k_y,0 |\; \widehat{T}^{2B}(j=0,K_z=0;z=0) \;
                           | 0,k_y',0 \rangle \nonumber\\
&& = \frac{-V_0}{L_y L_z}
     \phi^r_{0}(k_y l_r^2) \phi^r_{0}(k'_y l_r^2) \nonumber\\
&&   \times
            \left[ 1 + \frac{-V_0}{4\pi l_B^2 L_z} \sum_{N'',M'',k_z''}
                       \frac{|B_0^{N''M''}|^2}
                            {\varepsilon_{N'',k_z''}+\varepsilon_{M'',-k_z''}}
            \right]^{-1}~.
\end{eqnarray}
To extract an expression for the scattering length we put the
above matrix element equal to the matrix element $ \langle
N+M-j,k_y,k_z |\; V_{\rm pp} \; | N'+M'-j,k_y',k_z' \rangle$ of
the pseudo-potential $V_{\rm pp} ({\bf r}) = 4 \pi a_{\rm sc}
\hbar^2 \delta ({\bf r}) / m $. From this we find that
\begin{equation}
\label{eq:scatlength}
\frac{m}{4\pi \hbar^2 a_{\rm sc}} =
-\frac{1}{V_0}
+ \frac{1}{4\pi l_B^2 L_z}
\sum_{N,M,k_z}
\frac{|B_0^{NM}|^2}{\varepsilon_{N,k_z}+\varepsilon_{M,-k_z}}~.
\end{equation}

\end{appendix}


\end{document}